\documentclass[11pt,a4paper]{article}
\usepackage[T1]{fontenc}
\usepackage{color}
\usepackage{graphicx}
\usepackage{bm}
\usepackage{hyperref}
\usepackage[utf8]{inputenc}
\usepackage{comment}
\hypersetup{
 colorlinks=true,%
 linkcolor=blue,
 citecolor=blue,
}

\usepackage{amsmath,amsfonts,amssymb}

\begin{document}
\begin{titlepage}
\begin{flushright}
    { IPMU18-0101}
\end{flushright}
\vspace{1cm}

\begin{flushleft}
    \textbf{\textsf{\huge Big Bang Nucleosynthesis Constraint on \\[0.5em]
    Baryonic Isocurvature Perturbations}}
\end{flushleft}

\vspace{2cm}

\noindent
\textbf{\large Keisuke Inomata}$^{a,b}$,
\textbf{\large Masahiro Kawasaki}$^{a,b}$,
\textbf{\large Alexander Kusenko}$^{b,c}$,
\textbf{\large Louis Yang}$^{b}$

\vspace{0.5cm}

\noindent
$^{a}$ICRR, The University of Tokyo, Kashiwa, Chiba 277-8582, Japan\\[0.3em]
$^{b}$Kavli IPMU (WPI), UTIAS, The University of Tokyo, Kashiwa, Chiba 277-8583, Japan\\[0.3em]
$^{c}$Department of Physics and Astronomy, University of California, Los Angeles, CA 90095-1547, USA

\vspace{0.3cm}
\noindent
{\small 
}

\vspace{3cm}

\abstract{
We study the effect of large baryonic isocurvature perturbations on the abundance of deuterium (D) synthesized in big bang nucleosynthesis (BBN). 
We found that large baryonic isocurvature perturbations existing at the BBN epoch ($T\sim 0.1$~MeV) change the D abundance by the second order effect, which, together with the recent precise D measurement, leads to a constraint on the amplitude of the power spectrum of the baryon isocurvature perturbations.  The obtained constraint on the amplitude is $\lesssim 0.016 \,(2\sigma)$ for scale $k^{-1} \gtrsim 0.0025$~pc.  This gives the most stringent one for $0.1\,\text{Mpc}^{-1} \lesssim k \lesssim 4\times 10^8\, \text{Mpc}^{-1}$. 
We apply the BBN constraint to the relaxation leptogenesis scenario, where large baryon isocurvature perturbations are produced in the last $N_\text{last}$ $e$-fold of inflation, and we obtain a constraint on $N_\text{last}$.
}

\end{titlepage}


\section{Introduction}

Light elements such as $^4$He and D are synthesized at the cosmic temperature $T$ around $0.1$~MeV. 
The abundances of light elements predicted by the big-bang nucleosynthesis (BBN) are in good agreement with those inferred from the observations, which has supported the standard hot big-bang cosmology since 1960's~(for review see, \cite{Iocco:2008va}).
The BBN is very sensitive to physical conditions at $T\simeq 1-0.01$~MeV and hence it is an excellent probe to the early universe.
For example, an extra radiation component existing $T\sim 1$~MeV changes the predicted abundances of $^4$He and D and could spoil the success of the BBN, which gives a stringent constraint on the extra radiation energy.

In modern cosmology it is believed that the hot universe is produced after inflation which is an accelerated expansion of the universe and solves several problems in the standard cosmology. 
One of the most important roles of inflation is a generation of density perturbations. 
During inflation light scalar fields, including the inflaton, acquire quantum fluctuations which become classical by the accelerated expansion and leads to density perturbations.
If only one scalar field ($=$ inflaton) is involved in generation of the density perturbations the produced perturbations are adiabatic and nearly scale invariant, which perfectly agrees with the observations of the CMB and large scale structures on large scales ($ \gtrsim \mathcal{O}(10)$~Mpc).
The amplitude of the power spectrum of the curvature perturbations is precisely determined as $\mathcal{P}_\zeta = 2.1\times 10^{-9}$ at the pivot scale $k=0.002~\text{Mpc}^{-1}$ by the CMB observations~\cite{Ade:2015xua}.

However, as for small scales we know little about the shape and amplitude of the power spectrum of the density perturbations and there are a few constraints on the curvature perturbations from the CMB $\mu$-distortion due to the Silk dumping~\cite{Chluba:2013dna} and overproduction of primordial black holes~\cite{Carr:2009jm}. 
The BBN also gives constraints on the amplitude of the curvature perturbations since they can affect $n/p$ and/or baryon-to-photon ratio through the second order effects and hence change the abundance of the light elements~\cite{Jeong:2014gna,Nakama:2014vla,Inomata:2016uip} .

Inflation produces not only the curvature (adiabatic) perturbations but also the isocurvature ones.
The isocurvature perturbations are produced when multiple scalar fields are involved in generation of the density perturbations.
In particular, when scalar fields play an important role in baryogenesis, baryonic isocurvature perturbations are generally produced.
A well-known example is the Affleck-Dine baryogenesis~\cite{Affleck:1984fy} where a scalar quark has a large field value during inflation and generates baryon number through dynamics after inflation. 
This baryogenesis scenario produces the baryonic isocurvature perturbations unless the scalar quark has a large mass in the phase direction~\cite{Enqvist:1998pf,Kasuya:2008xp}.
Since the CMB observations are quite consistent with the curvature perturbations, the isocurvature perturbations on the CMB scales are stringently constrained~\cite{Ade:2015lrj}.
However, there are almost no constraints on small-scale isocurvature perturbations.

In this paper we show that large baryonic isocurvature perturbations existing at the BBN epoch ($T\sim 0.1$~MeV) change the D abundance by their second order effect.
Because the primordial abundance of D is precisely measured with accuracy about 1~\%~\cite{Zavarygin:2018dbk,Cooke:2017cwo}, we can obtain a significant constraint on the amplitude $\mathcal{P}_{S_B}$ of the baryonic isocurvature perturbations. 
It is found that the amplitude should be $\mathcal{P}_{S_B} \lesssim 0.016 \,(2\sigma)$ for scale $k^{-1} \gtrsim 0.0025$~pc. 
We also apply the constraint to the relaxation leptogenesis scenario~\cite{Kusenko:2014lra,Yang:2015ida} where large fluctuations of a scalar field play a crucial role in leptogenesis and large baryonic isocurvature perturbations are predicted.
We show that the BBN gives a significant constraint on this scenario. 

The paper is organized as follows.
In Sec.~\ref{sec:introduction} we briefly review the measurement of the D abundance.
We show how the baryonic isocurvature perturbations change the D abundance and obtain a generic constraint on their amplitude in Sec.~\ref{sec:baryon_isocurvature}. 
In Sec.~\ref{sec:relaxation_leptogenesis} we apply the BBN constraint on the relaxation leptogenesis scenario.
Sec.~\ref{sec:conclusion} is devoted for conclusions.

\section{Deuterium abundance}
\label{sec:introduction}

\begin{figure}
  \centering \includegraphics[width=11cm]{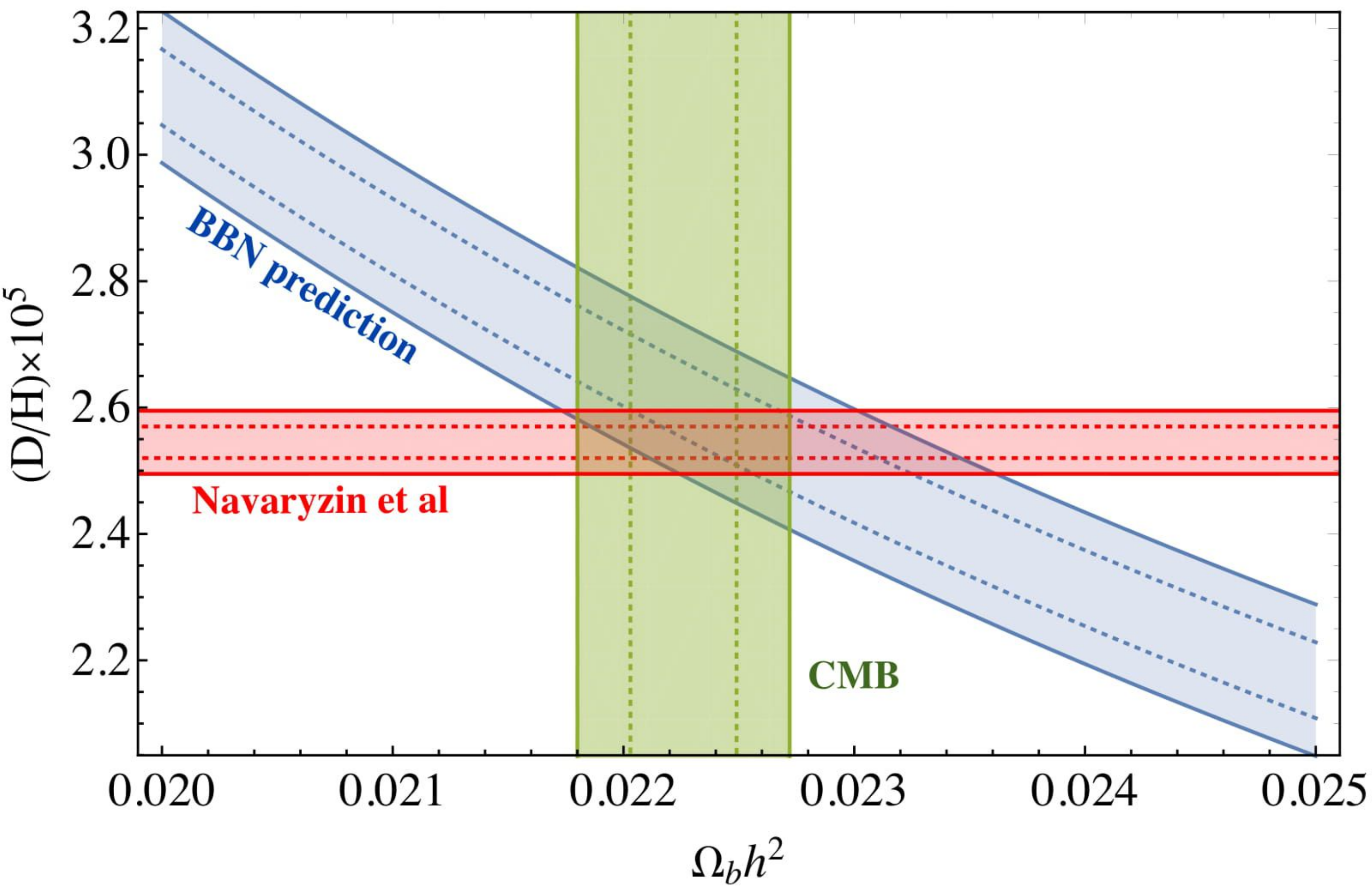}
  \caption{
  BBN prediction of deuterium as a function of the baryon  density is shown by the blue band.
  The CMB constraint on the baryon density (green) and the observed D abundance (red) are also shown. The dotted (solid) lines denote the contours of $1\sigma$ ($2\sigma$) uncertainties.
}
  \label{fig:deuterium_2sigma}
\end{figure}

Light elements like D, $^3$He and $^4$He are synthesized in the early Universe at temperature $T\simeq 1~\text{MeV}-0.01~\text{MeV}$.
This big bang nucleosynthesis predicts the abundances of light elements which are in agreement with those inferred by observations.
In particular, the deuterium abundance has been precisely measured by observing absorption of QSO lights due to damped Lyman-$\alpha$ systems.
Most recently Zavaryzin et al.~\cite{Zavarygin:2018dbk} reported the primordial D abundance,
\begin{equation}
    (\text{D}/\text{H})_p = (2.545 \pm 0.025)\times 10^{-5},
    \label{eq:D_obs}
\end{equation}
from measurements of 13 damped Lyman-$\alpha$ systems.
Here $\text{D}/\text{H}$ is the ratio of the number densities of D and H.
The observed abundance should be compared with the theoretical prediction.
The D abundance produced in BBN is calculated by numerically solving the nuclear reaction network and in the standard case the result is only dependent on the baryon density $\Omega_B$.
We adopt the following fitting formula in Ref.~\cite{Ade:2015xua}:
\begin{equation}
    10^5 (\text{D}/\text{H})_p = 18.754-1534.4\,\omega_B 
    + 48656\,\omega_B^2
    -552670\,\omega_B^3,
    \label{eq:D_prediction}
\end{equation}
where $\omega_B = \Omega_B h^2$ and $h$ is the Hubble constant in units of 100km/s/Mpc.
This formula is obtained using the PArthENoPE code~\cite{Pisanti:2007hk} and its uncertainty is $\pm 0.12 (2\sigma)$.
The observational constraint Eq.~(\ref{eq:D_obs}) and the prediction Eq.~(\ref{eq:D_prediction}) are shown in Fig.~\ref{fig:deuterium_2sigma}.
From the figure the BBN prediction is consistent with the observed abundance for $\Omega_B h^2 \simeq 0.022-0.023$.

The baryon density is also precisely determined by CMB observations.
The recent Plank measurement gives
\begin{equation}
    \Omega_B h^2 = 0.02226\pm 0.00023,
\end{equation}
which is also shown in Fig.~\ref{fig:deuterium_2sigma}.
It is seen that the baryon densities determined by BBN and CMB are consistent.
However, if the predicted D abundance increases by about 3\% in the case with $2\sigma$ uncertainties, 
 they become inconsistent and hence any effect that increases the D abundance is stringently constrained.

\section{Baryonic isocurvature perturbations and deuterium abundance}
\label{sec:baryon_isocurvature}

Here we assume that only baryon number fluctuations are produced in the early universe.
Such fluctuations are called baryonic isocurvature density perturbations $S_B$ which are written as
\begin{equation}
    S_B = \frac{\delta n_B}{n_B}
    -\frac{3}{4}\frac{\delta\rho_\gamma}{\rho_\gamma}
    =\frac{\delta n_B}{n_B},
\end{equation}
where $\rho_\gamma$ and $\delta\rho_\gamma$ are the photon energy density and its perturbation, and we have used the above assumption of the nonexistence of photon perturbations in the last equality.

\begin{figure}
  \centering \includegraphics[width=11cm]{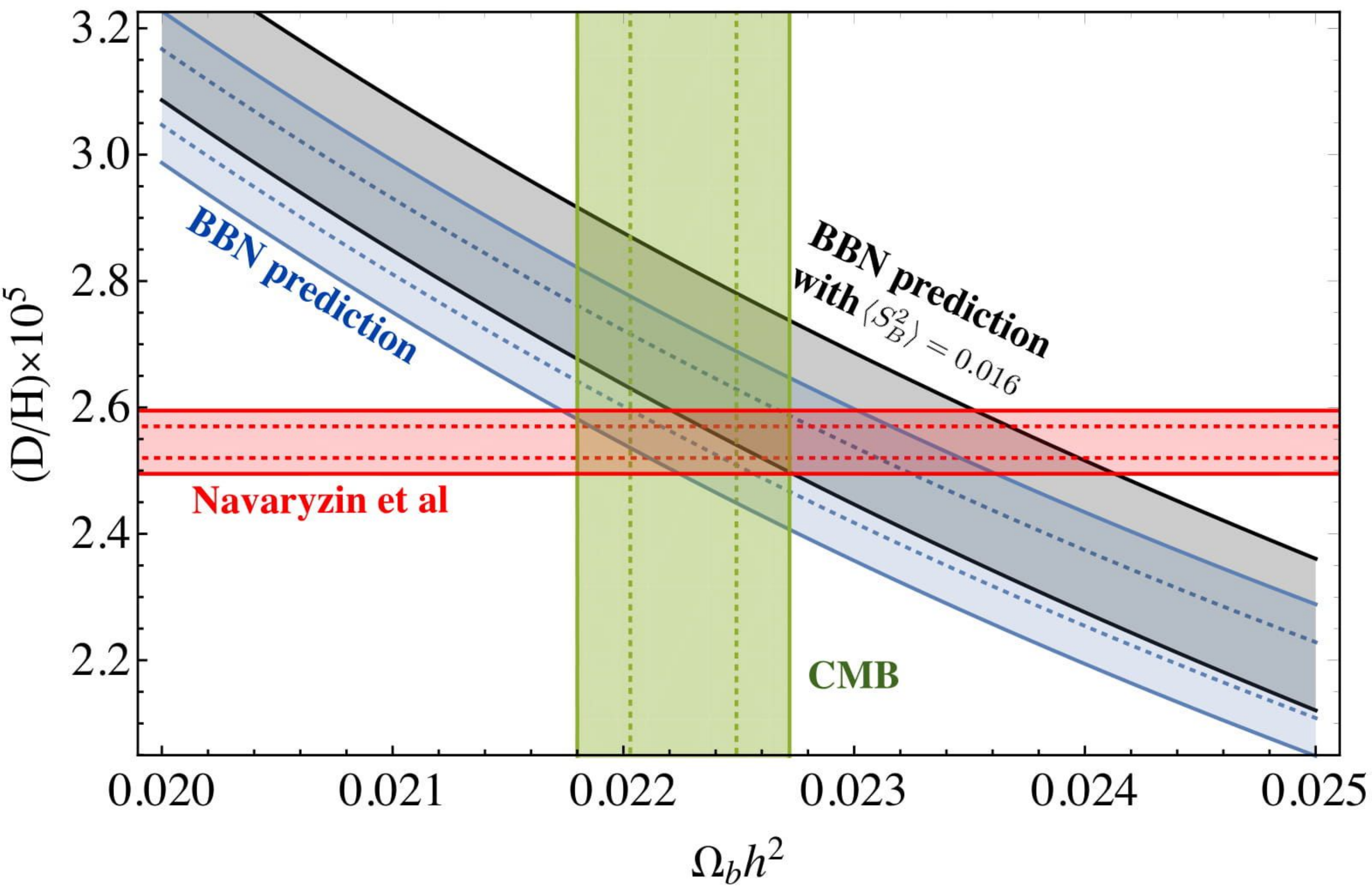}
  \caption{
	The same as Fig.~\ref{fig:deuterium_2sigma}, except that we show the $2 \sigma$ BBN prediction in the case with $\langle S_B^2 \rangle =0.016$ by the gray shaded region.
    Note that the overlap between the region of the D observation and BBN prediction does not necessarily mean the consistence between them.
	}
  \label{fig:deuterium_S01_2sigma}
\end{figure}

When the baryon number density has spatial fluctuations it can affect the BBN and change the abundance of light elements.
In particular, modification of the D abundance is important because it is precisely determined by the recent measurement.
Let us consider the BBN prediction of D in the presence of the baryon number fluctuations by using Eq.~(\ref{eq:D_prediction}).
In order to take into account the baryon number fluctuations, we consider $\omega_B$  in Eq.~(\ref{eq:D_prediction}) as space dependent variable,
\begin{equation}
    \omega_B(t,\vec{x}) = \bar{\omega}_B + \delta\omega_B(\vec{x}),
\end{equation}
where $\bar{\omega}_B$ is the homogeneous part and $\delta\omega_B$ denotes the fluctuations which are related to $S_B$ as
\begin{equation}
    \delta\omega_B = \bar{\omega}_B \frac{\delta n_B(\vec{x})}{n_B}
    =\bar{\omega}_B S_B(\vec{x}).
\end{equation}
With $\bar{\omega}_B$ and $S_B$, Eq.~(\ref{eq:D_prediction}) is rewritten as
\begin{align}
    y_d = & 18.754-1534.4\,\bar{\omega}_B + 
         48656\,\bar{\omega}_B^2
           -552670\,\bar{\omega}_B^3 \nonumber \\
         & + (-1534.4 \,\bar{\omega}_B  + 97312\,\bar{\omega}_B ^2
           - 1658010\,\bar{\omega}_B^3)\,S_B \nonumber \\
         & +(48656\,\bar{\omega}_B ^2 - 1658010\,\bar{\omega}_B ^3)\,S_B^2
           +\ldots,
\end{align}
where $y_d = 10^5(\text{D}/\text{H})_p$.
Since D production takes place at $T\simeq 0.1$~MeV we estimate $S_B$ at that time.
To estimate the primordial D abundance we should average $y_d$ over the volume $V$ corresponding to the present horizon.
Using $\langle S_B\rangle =0$, we obtain
\begin{align}
    \langle y_d \rangle = & 18.754-1534.4\,\bar{\omega}_B 
    + 48656\,\bar{\omega}_B^2
           -552670\,\bar{\omega}_B^3 \nonumber \\
         & +(48656\,\bar{\omega}_B ^2 - 1658010\,\bar{\omega}_B ^3)\,
           \langle S_B^2\rangle,
\label{eq:yd_formula_ave}           
\end{align}
where $\langle \cdots \rangle$ denotes the spatial average.
Thus, the D abundance is modified from the homogeneous case owing to the second order effect of the baryonic isocurvature perturbations.
The prediction for $\langle S_B^2 \rangle = 0.016$ is shown in Fig.~\ref{fig:deuterium_S01_2sigma}.
It is seen that the isocurvature perturbations increase the D abundance and hence increase the baryon density accounting for the observed abundance, which leads to inconsistency between the baryon densities inferred from CMB and BBN.
Thus we can obtain a constraint on $\langle S_B^2 \rangle$.

In order to derive the upper bound on the isocurvature perturbations, 
we define the discrepancy $\mathcal{D}$ between the observational and theoretical values in the units of standard deviation as 
\begin{align}
\mathcal{D} \equiv 
\frac{|y_\text{obs,mean} - y_\text{th,mean}|}{\sqrt{\sigma^2_{y_\text{obs}} + \sigma^2_{y_\text{th}}}},
\label{eq:disc_formula}           
\end{align}
where $y_\text{obs,mean}$ and $y_\text{th,mean}$ are the mean values of the observation and theoretical prediction 
and $\sigma^2_\text{obs}$ and $\sigma^2_\text{th}$ are the standard deviations of $y_\text{obs}$ and $y_\text{th}$.
Note that $y_\text{th,mean}$ and $\sigma_{y_\text{th}}$ are calculated by Eq.~(\ref{eq:yd_formula_ave}) and therefore they depend on $\langle S_{B}^2 \rangle$.
Imposing the conditions, $\mathcal{D}<1 $ or $\mathcal{D}<2$, we can get the constraints on the isocurvature perturbations as $\langle S_B^2 \rangle < 0.0020$ ($1\sigma$) or $\langle S_B^2 \rangle < 0.016$ ($2\sigma$).

Let us calculate $\langle S_B^2 \rangle$ from the Fourier mode $S_B(\vec{k})$ as
\begin{align}
    \langle S_B^2\rangle & = \frac{1}{V}\int_V d^3 x (S_B(\vec{x}))^2\nonumber \\[0.5em]
     & = \frac{1}{(2\pi)^6V}\int_V d^3x
    \int d^3k \int d^3k' S_B(\vec{k})S_B^*(\vec{k'})e^{i(\vec{k}-\vec{k'})\cdot \vec{x}}
    \nonumber \\[0.5em]
    & = \frac{1}{(2\pi)^3}\int d^3 k |S_B(\vec{k})|^2
    = \int d\ln k \,\mathcal{P}_{S_B}(k),
\end{align}
where $\mathcal{P}_{S_B}(k)$ is the power spectrum of the baryonic isocurvature perturbations.
Here we should pay attention to the upper limit of the $k$-integration.
The baryons diffuse in the early universe, which erases the baryon number fluctuations with their wavelength less than the diffusion length.
So the upper limit of the integration is the wave number $k_d$ which corresponds to the diffusion length at the BBN epoch ($T\simeq 0.1$~MeV).
The diffusion length of neutrons $d_n$ is much larger than that of protons, so $k_d$ is given by $d_n^{-1}$.
The neutron diffusion is determined by neutron-proton scatterings and the comoving diffusion length is given by~\cite{Applegate:1987hm}
\begin{equation}
    d_n \simeq k_d^{-1} \simeq 0.0025\, \text{pc}~~~~~\text{at }~T=0.1\,\text{MeV}.
\end{equation}
Thus, $\langle S_B^2 \rangle$ is given by
\begin{equation}
    \langle S_B^2\rangle = \frac{1}{(2\pi)^3}\int^{k_d}_{k_*} d^3k P_{S_B}(k).
    \label{eq:isocuravture_fluc}
\end{equation}
Here $k_*$ is the scale corresponding to the present horizon ($k_*^{-1} \simeq 3000h^{-1}$~Mpc).

Here, let us summarize the constraints on power spectra of baryonic isocurvature perturbations.
In addition to the BBN constraint, which we have discussed so far, there are constraints from the observations of the CMB anisotropy and large scale structure (LSS)~\cite{Ade:2015lrj,Seljak:2006bg,Beltran:2005gr}.\footnote{
Large isocurvature perturbations could possibly make the CMB distortion.
However, the produced distortion is too small to constrain the power spectrum with the current observations~\cite{Chluba:2013dna}.}
From the CMB anisotropy observations, the effective cold dark matter (CDM) isocurvature perturbations are constrained as (95\% CL)~\cite{Ade:2015lrj}
\begin{align}
      \begin{cases}
    \beta_\text{iso,CDM}(k_\text{low}) < 0.045 & (k_\text{low}=0.002 \, \text{Mpc}^{-1} ) \\
    \beta_\text{iso,CDM}(k_\text{mid}) < 0.379 & (k_\text{mid}=0.05 \, \text{Mpc}^{-1} )\\
    \beta_\text{iso,CDM}(k_\text{high}) < 0.594 & (k_\text{high}=0.1 \, \text{Mpc}^{-1} ).
    \label{eq:cmb_b_iso_cons}
  \end{cases}
\end{align}
$\beta_\text{iso,CDM}(k)$ is defined as
\begin{equation}
	\beta_\text{iso,CDM}(k) \equiv \frac{\mathcal P_{S_\text{CDM,eff}}(k)}{\mathcal P_{\zeta}(k) + \mathcal P_{S_\text{CDM,eff}}(k)},
    \label{eq:cdm_iso_cons}
\end{equation}
where $\mathcal P_{S_\text{CDM,eff}}$ is the power spectrum of the effective CDM isocurvature perturbations, which is defined as $S_\text{CDM,eff} = S_\text{CDM} + \frac{\Omega_B h^2}{\Omega_\text{CDM} h^2} S_\text{B}$ with the CDM energy density parameter $\Omega_\text{CDM}h^2 (=0.119)$. 
In the case with baryonic isocurvature perturbations but without CDM ones, the relation $\mathcal P_{S_\text{CDM,eff}} = \left(\frac{\Omega_B h^2}{\Omega_\text{CDM} h^2}\right)^2 \mathcal P_{S_B}$ is satisfied.
Then we can convert the constraints on $\mathcal P_{S_\text{CDM,eff}}$ into those on $\mathcal P_{S_B}$.
Meanwhile, from the combination of the LSS, such as the Lyman-$\alpha$ forest anisotropy, and CMB observations, the isocurvature perturbations are constrained as (95\% CL)~\cite{Seljak:2006bg}
\begin{align}
	A_\text{iso,bar} = -0.06^{+0.35}_{-0.34},
    \label{eq:lya_b_iso_cons}	
\end{align}
where $A^2_\text{iso,bar} \equiv \mathcal P_{S_B}(k_0) / \mathcal P_{\zeta}(k_0)$ ($k_0 = 0.05$\,Mpc$^{-1}$) 
and this constraint is based on the assumptions that the spectral index of the isocurvature perturbations is the same as that of the curvature perturbations and the isocurvature perturbations are fully correlated with the curvature perturbations.
A positive value of $A_\text{iso,bar}$ means the full positive correlation and a negative one means the full negative (or anti-) correlation.
Note that the Lyman-$\alpha$ forest observations can see the power spectra in smaller scale ($k\lesssim 1$\,Mpc$^{-1}$) than the CMB observations can ($k\lesssim 0.1$\,Mpc$^{-1}$).

\begin{figure}
  \centering \includegraphics[width=11cm]{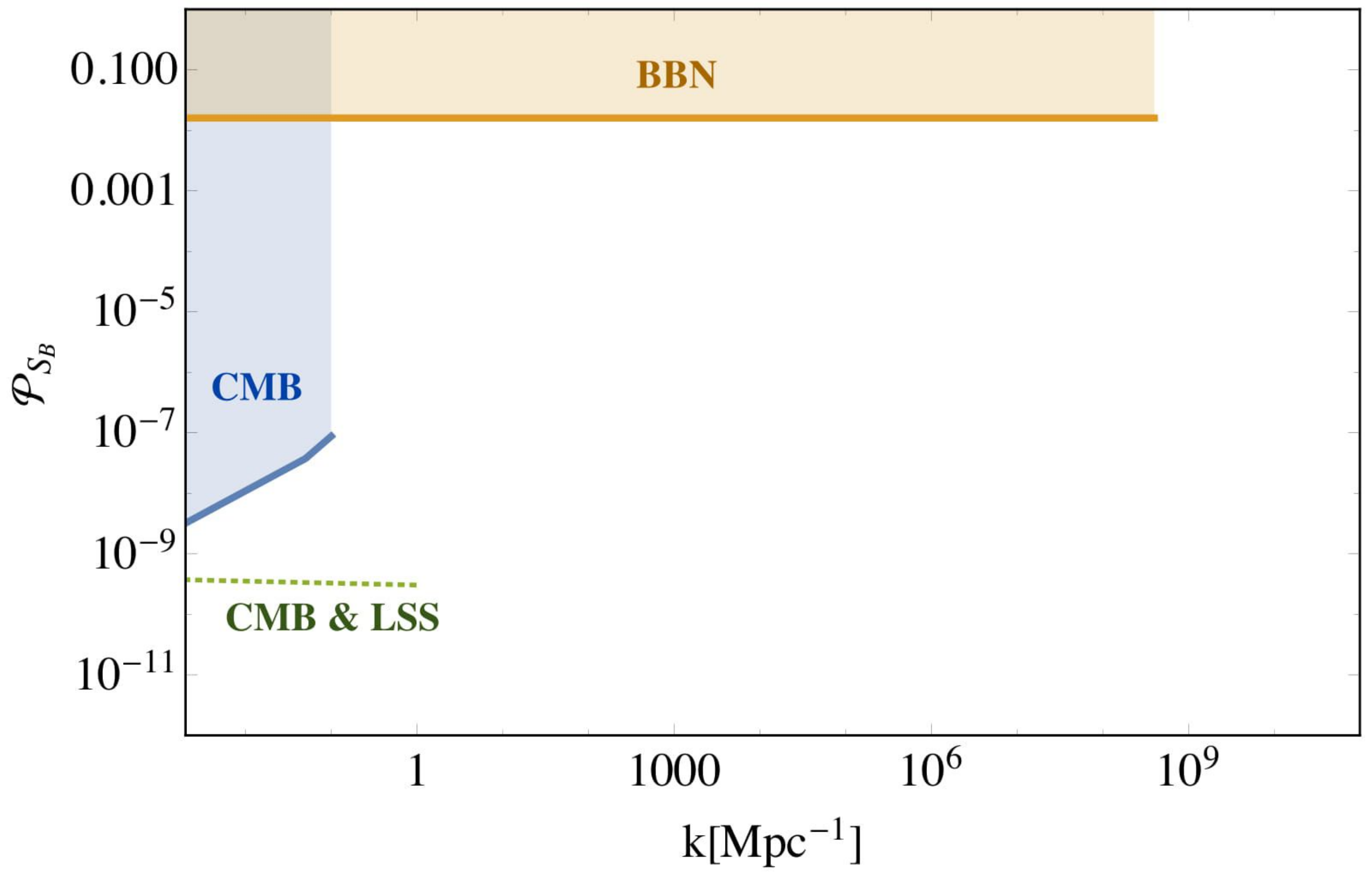}
  \caption{
	The summary of the constraints on baryonic isocurvature perturbations.
    An orange shaded region is excluded by the D observations, which we derived in this paper.
	A blue shaded region is excluded by the CMB observations~\cite{Ade:2015lrj}. 
    For comparison, we plot also the constraints from the combination of CMB and LSS observations with a green dotted line~\cite{Seljak:2006bg},
    though this constraint is based on some assumptions (see text).
	}
  \label{fig:b_iso_const}
\end{figure}

To visualize the BBN constraints, we assume that the power spectrum is monochromatic as
\begin{align}
	\mathcal P_{S_B,\text{mono}}(k;k_*) = \mathcal P_{S_B} \delta(\text{log} k - \text{log} k_* ).
\end{align}
With this monochromatic power spectrum, the equation $\langle S_B^2\rangle = \mathcal P_{S_B}$ is satisfied.
In Fig.~\ref{fig:b_iso_const}, we show the $\mathcal P_{S_B}$ region excluded by the BBN observations with orange shaded one.
For comparison, we show also the constraints from the CMB and LSS observations given by Eqs.~(\ref{eq:cmb_b_iso_cons}) and (\ref{eq:lya_b_iso_cons}). 
Regarding the combination constraint in Fig.~\ref{fig:b_iso_const}, to show the conservative constraint, we assume the full negative correlation and take $A_\text{iso,bar} = -0.40$.
Note that the derived constraint on $\langle S_B^2 \rangle$ is valid even in the compensated isocurvature perturbations~\cite{Gordon:2002gv}, in which the CDM isocurvature perturbation totally compensates the baryonic one, because BBN occurs during radiation era and is independent of CDM perturbations.\footnote{
Compensated isocurvature perturbations on large scale can be constrained from CMB and baryon acoustic oscillation observations~\cite{Grin:2013uya,Munoz:2015fdv,Valiviita:2017fbx,Haga:2018pdl,Soumagnac:2016bjk,Soumagnac:2018atx}.
}

Before closing this section, we discuss the applicability of Eq.~(\ref{eq:D_prediction}) to the inhomogeneous BBN.
Since D is synthesized in a rather short period ($T=0.1$--$0.05$~MeV), we can focus on only that period.
The most important scale in our problem is the diffusion length. 
In Eq.~(\ref{eq:isocuravture_fluc}) we take the neutron diffusion length $d_n$ as a cutoff because the proton diffusion length $d_p$ is about $100$ times shorter than $d_n$~\cite{Applegate:1987hm}. 
If the fluctuation wavelength $k^{-1}$ of baryons are larger than the neutron diffusion scale $d_n$, the diffusion does not play an important role and D production takes place in locally homogeneous regions, which justifies the use of Eq.~(\ref{eq:D_prediction}). 
Thus, our generic constraint in Fig.~\ref{fig:b_iso_const} is valid for $k < k_d=d_n^{-1}$.
On the other hand, if $k^{-1}$ is smaller than the proton diffusion length, diffusion makes baryons homogeneous, so again we can use Eq.~(\ref{eq:D_prediction}).
The complicated situation occurs when the fluctuation scale is $d_n^{-1} < k < d_p^{-1}$.
To see how BBN is affected by such fluctuations, let us consider a region with size $d_n$. 
In this region neutrons are homogeneous but protons fluctuate, which produces high and low proton sub-regions in a homogeneous neutron region.
In high proton sub-regions BBN proceeds a little earlier than in low proton sub-regions, which
leads to more $^4$He and hence less D.
Because the fluctuations are perturbative,  the net result is the same as the homogeneous BBN in the first order. 
However, if we take the second order effect into account, we have to consider the diffuse-back of neutrons.
In the high proton sub-regions neutron are consumed for D production earlier and neutron density 
becomes smaller than the low proton sub-regions. 
Then neutrons in the low proton sub-region diffuse into the high proton sub-regions and those neutrons are consumed. 
Thus, as a net result, D could decrease by the second order effect. 
Since it is difficult to estimate this effect quantitatively, we assume that the effect is small in this paper.

\section{Relaxation leptogenesis}
\label{sec:relaxation_leptogenesis}
 
One model capable of generating the baryonic isocurvature perturbations is the relaxation leptogenesis~\cite{Kusenko:2014lra,Pearce:2015nga,Yang:2015ida}. In this section, we will briefly review the relaxation leptogenesis framework and discuss the improved constraint from the deuterium abundance on this type of models. In the relaxation leptogenesis framework, the generation of lepton/baryon asymmetry is driven by the classical motion of a scalar field $\phi$ during the period of cosmic inflation and the following reheating stage of the universe. During inflation, a light scalar field $\phi$ with mass $m_{\phi}<H_{I}$ can develop a large nonzero vacuum expectation value (VEV) $\phi_{0}\equiv\sqrt{\left\langle \phi^{2}\right\rangle }$ through quantum fluctuations~\cite{Linde:1982uu,Starobinsky:1982ee,Vilenkin:1982wt}. If the quantum fluctuation is not suppressed by the potential or other interactions, the $\phi$ can reach an equilibrium VEV $\phi_{0}$ satisfying $V\left(\phi_{0}\right)\sim H_{I}^{4}$, where $H_{I}=\Lambda_{I}^{2}/\sqrt{3}M_{pl}$ is the Hubble rate during inflation, and $\Lambda_{I}$ is the inflationary energy scale.

However, in general, one can expect there are interactions between $\phi$ and the inflaton field $I$ of the form 
\begin{equation}
\mathcal{L}_{\phi I}=\lambda_{\phi I}\frac{\left(\phi^{\dagger}\phi\right)^{m/2}\left(I^{\dagger}I\right)^{n/2}}{M_{pl}^{m+n-4}}.\label{eq:inflaton Higgs couplings}
\end{equation}
In the early stage of the inflation when the inflaton VEV $\left\langle I\right\rangle $ is large, interactions like Eq.~\eqref{eq:inflaton Higgs couplings} can contribute a large effective mass term {[}$m_{\phi}\left(I\right)\gg H_{I}${]} to $\phi$ suppressing the quantum fluctuations of $\phi$. As the inflaton VEV $\left\langle I\right\rangle $ decreases in the later stage of inflation, $\phi$ becomes lighter. When the effective mass of $\phi$ falls below $m_{\phi}\left(I\right)<H_{I}$, the quantum fluctuations of $\phi$ can start to grow. If the VEV of $\phi$ only develops in the last $N_{\mathrm{last}}$ $e$-folds of inflation, its VEV can reach $\phi_{0}\simeq\sqrt{N_{\mathrm{last}}}H_{I}/2\pi$. This is the ``IC-2'' scenario considered in \cite{Kusenko:2014lra,Yang:2015ida}.

During reheating, the VEV of $\phi$ relaxes to the minimum of the potential and oscillates with decreasing amplitudes. The relaxation of $\phi$ provides the out of thermal equilibrium condition and breaks time-reversal symmetry, allowing baryogenesis to proceed. For successful relaxation leptogenesis, one considers the derivative coupling between the $\phi$ and the $B+L$ fermion current $j_{B+L}^{\mu}$ of the form 
\begin{equation}
\mathcal{O}_{6}=-\frac{1}{\Lambda_{n}^{2}}\left(\partial_{\mu}\left|\phi\right|^{2}\right)j_{B+L}^{\mu},
\end{equation}
for some higher energy scale $\Lambda_{n}$. This operator can be treated as an effective chemical potential for the fermion current $j_{B+L}$ as $\phi$ evolves in time. In the presence of a $B$ or $L$-violating process, the system can then relax toward a state with nonzero $B$ or $L$.

In the case of Higgs relaxation leptogenesis ($\phi=h$), the final lepton asymmetry, $Y\equiv n_{L}/s$, is estimated to be \cite{Kusenko:2017kdr} 
\begin{equation}
Y\approx\dfrac{90\sigma_{R}}{\pi^{6}g_{*S}}\left(\dfrac{\phi_{0}}{\Lambda_{n}}\right)^{2}\dfrac{3z_{0}T_{RH}}{4\alpha_{T}t_{RH}}\exp\left(-\dfrac{8+\sqrt{15}}{\pi^{2}}\sigma_{R}T_{RH}^{3}t_{RH}\right),\label{eq:Y_approx_final}
\end{equation}
if the Higgs potential is dominated by the thermal mass term $V\left(\phi,T\right)\approx\frac{1}{2}\alpha_{T}^{2}T^{2}\phi^{2}$ during reheating. The parameters for Eq.~\eqref{eq:Y_approx_final} are $\alpha_{T}\approx0.33$ at the energy scale $\mu\sim10^{13}$ GeV, $g_{*S}=106.75$, and $z_{0}=3.376$. $\sigma_{R}$ is the thermally averaged cross section of the $L$-violating process, which we consider to be the scattering between left-handed neutrinos via the exchange of a heavy right-handed neutrino. $\phi_{0}$ is the initial VEV of the Higgs field at the end of inflation, which depends on $N_{\mathrm{last}}$. The reheating channel is assumed to be perturbative with the reheat temperature $T_{RH}\simeq\left(24/\pi^{2}g_{*}\right)^{1/4}\sqrt{M_{pl}/t_{RH}}$ when reheating is complete at $t_{RH}$. The produced lepton asymmetry then turns into baryon asymmetry through the Sphaleron process and becomes the baryon density of the universe as the universe cools down.

Since the asymmetry generated in this manner depends crucially on the initial VEV $\phi_{0}$, the spatial fluctuation of $\phi$ due to quantum fluctuation at inflation stage can result in baryonic perturbations at later time. These perturbations are isocurvature modes because the scalar field $\phi$ is not the inflaton $I$ and doesn't dominate the energy density of the universe. As we have discussed in previous sections, baryonic isocurvature perturbations are constrained by observations in both the amplitude and the spatial scale $k$. Thus, we can translate these into the constraints on the $N_{\mathrm{last}}$ of the Higgs field and other inflation parameters like $\Lambda_{I}$ and $T_{RH}$.

As computed in Ref.~\cite{Kusenko:2017kdr}, the power spectrum of the baryon density perturbations resulted from the quantum fluctuation of $\phi$ is 
\begin{equation}
\mathcal{P}_{S_{B}}\left(k\right)\approx\frac{4}{N_{\mathrm{last}}^{2}}\ln\left(\frac{k}{k_{s}}\right)\theta\left(k-k_{s}\right)\theta\left(k_{s}e^{N_{\mathrm{last}}}-k\right),
\end{equation}
for the fluctuation of $\phi$ only developed in the last $N_{\mathrm{last}}$ $e$-fold of inflation. Here $k_{s}\sim a\left(N_{\mathrm{last}}\right)H_{I}$ is the comoving wave number corresponding to the mode which first leaves the horizon. By considering a typical inflation setup with the inflationary energy scale $\Lambda_{I}$ and the reheat temperature $T_{RH}$, we can then relate $k_{s}$ and $N_{\mathrm{last}}$ by 
\begin{equation}
k_{s}\simeq2\pi e^{-N_{\mathrm{last}}}H_{I}\left(\frac{T_{RH}}{\Lambda_{I}}\right)^{4/3}\frac{g_{*S}^{1/3}\left(T_{\mathrm{now}}\right)}{g_{*S}^{1/3}\left(T_{RH}\right)}\frac{T_{\mathrm{now}}}{T_{RH}},
\end{equation}
or 
\begin{equation}
k_{s}e^{N_{\mathrm{last}}}\simeq65\,\mathrm{Mpc}^{-1}e^{46.3}\equiv k_{s,\,0}\,e^{N_{0}},\label{eq:ks_Nlast}
\end{equation}
for $\Lambda_{I}=10^{16}\,\mathrm{GeV}$, $T_{RH}=10^{12}\,\mathrm{GeV}$, and $T_{\mathrm{now}}=2.726\,\mathrm{K}$.

The square of the baryonic isocurvature perturbations $\left\langle S_{B}^{2}\right\rangle $ is then given by 
\begin{align}
\left\langle S_{B}^{2}\right\rangle  & =\int_{k_{*}}^{k_{d}}\frac{dk}{k}\mathcal{P}_{S_{B}}\left(k\right)\\
 & \approx\frac{4}{N_{\mathrm{last}}^{2}}\int_{k_{*}}^{k_{d}}\frac{dk}{k}\ln\left(\frac{k}{k_{s}}\right)\theta\left(k-k_{s}\right)\theta\left(k_{s}e^{N_{\mathrm{last}}}-k\right)\\
 & =\frac{2}{N_{\mathrm{last}}^{2}}\min\left[\ln^{2}\left(\frac{k_{d}}{k_{s}}\right)\theta\left(k_{d}-k_{s}\right),\,N_{\mathrm{last}}^{2}\right],
\end{align}
where in the last step we have assumed $k_{*}<k_{s}$. With Eq.~\eqref{eq:ks_Nlast}, we have 
\begin{equation}
\left\langle S_{B}^{2}\right\rangle =\frac{2}{N_{\mathrm{last}}^{2}}\left[\ln\left(\frac{k_{d}}{k_{s,\,0}}\right)+N_{\mathrm{last}}-N_{0}\right]^{2}\theta\left(k_{d}-k_{s}\right).
\end{equation}
The constraint on baryonic isocurvature perturbation from the D abundance then gives an upper bound on $N_{\mathrm{last}}$ as 
\begin{equation}
N_{\mathrm{last}}\lesssim33.7\quad\text{(}2\sigma\text{)},
\label{eq:N_const}
\end{equation}
for $\Lambda_{I}=10^{16}\,\mathrm{GeV}$ and $T_{RH}=10^{12}\,\mathrm{GeV}$.

\begin{figure}
\begin{centering}
\includegraphics[width=11cm]{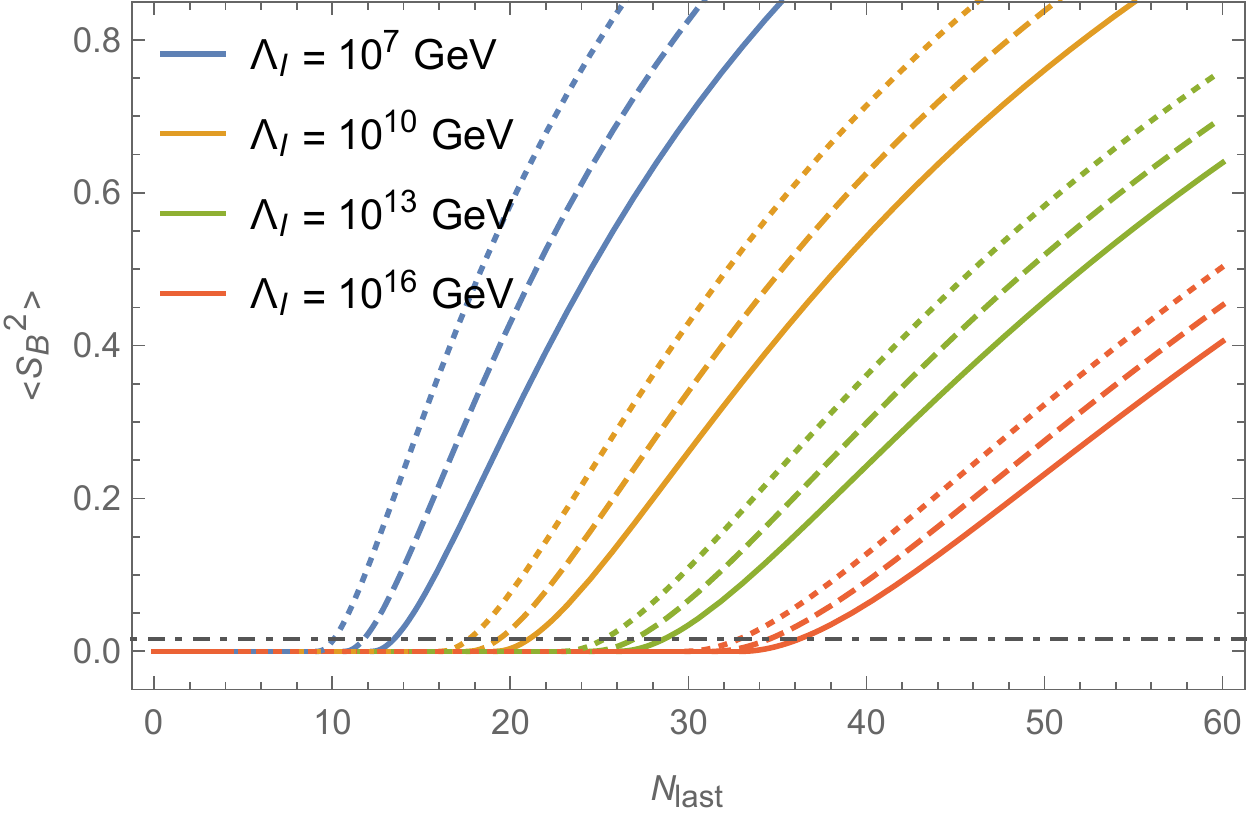} 
\par\end{centering}
\caption{The baryonic isocurvature perturbations $\left\langle S_{B}^{2}\right\rangle $ generated by the Higgs relaxation leptogenesis at various $N_{\mathrm{last}}$ and $\Lambda_{I}$. The solid, dashed, and dotted lines correspond to the cases where the reheat temperatures $T_{RH}$ are $10^{-1}\Lambda_{I}$, $10^{-3}\Lambda_{I}$, and $10^{-5}\Lambda_{I}$, respectively. The gray horizontal dash-dot line at $\langle S_B^2 \rangle = 0.016$ indicates the constraint from the D abundance at $2\sigma$ level. \label{fig:SB2 vs Nlast}}
\end{figure}

Figure \ref{fig:SB2 vs Nlast} shows the baryonic isocurvature perturbations generated by the Higgs relaxation leptogenesis model at various $N_{\mathrm{last}}$ and $\Lambda_{I}$. We see that larger values of $\Lambda_{I}$ and $T_{RH}$ allowing for larger $N_{\mathrm{last}}$. We also see that for each choice of parameters ($\Lambda_{I}$, $T_{RH}$), there is a minimum $N_{\mathrm{last}}$ below which the fluctuation $\left\langle S_{B}^{2}\right\rangle $ becomes zero. This corresponds to the case when the scale of the produced baryonic perturbations is smaller than the baryon diffusion scale $k_{d}$. So the baryonic perturbation is washed out by neutron diffusion before the BBN.

\begin{figure}
\begin{centering}
\includegraphics[height=8cm]{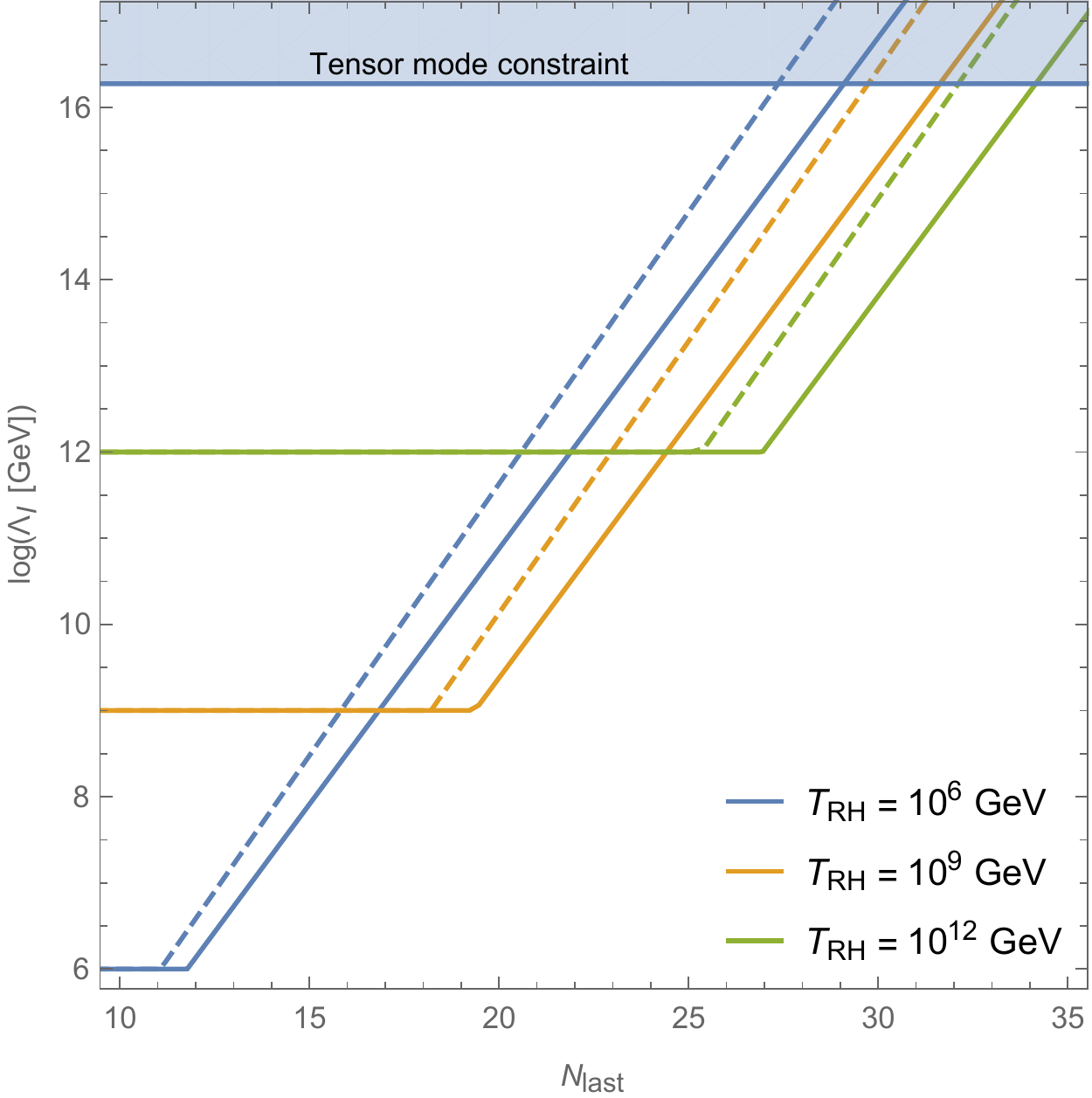} 
\par\end{centering}
\caption{The BBN constraint on the Higgs relaxation leptogenesis in the $\Lambda_{I}$-$N_{\mathrm{last}}$ space at various reheat temperature $T_{RH}$. The dashed (solid) lines correspond to the 1$\sigma$ (2$\sigma$) constraints. The regions on the lower right side of each contours are excluded. The blue shaded region for $\Lambda_{I} > 1.88\times10^{16}\,\text{GeV}$ is constrained by Planck non-observation of tensor mode~\cite{Ade:2015lrj}. For successful inflation, one also requires $\Lambda_{I} > T_{RH}$, which is indicated as the horizontal parts of each contours.  \label{fig:Lambda_I_vs_Nlast}
}
\end{figure}

Figure \ref{fig:Lambda_I_vs_Nlast} shows the parameter space in $\Lambda_{I}$ vs $N_{\mathrm{last}}$ at various reheat temperatures $T_{RH}$. Note that the CMB observations from Planck gives an upper bound on the inflationary energy scale $\Lambda_{I}<1.88\times10^{16}\,\text{GeV}$~\cite{Ade:2015lrj}. For successful Thus, for a given set of $\Lambda_{I}$ and $T_{RH}$, the D abundance constraint provides an upper bound on $N_{\mathrm{last}}$.

\section{Conclusion}
\label{sec:conclusion}

We have shown that large baryonic isocurvature perturbations existing at the BBN epoch ($T\sim 0.1$~MeV) change the D abundance by the second order effect, which, together with the recent precise D measurement with accuracy about $1$~\%, leads to a constraint on the amplitude of the power spectrum $\mathcal{P}_{S_B}$ of the baryon isocurvature perturbations. 
It is found that the amplitude should be $\mathcal{P}_{S_B} \lesssim 0.016 \,(2\sigma)$ for scale $k^{-1} \gtrsim 0.0025$~pc [see Fig.~\ref{fig:b_iso_const}].  
Since there has been no constraint on baryonic isocurvature perturbations on small scale $k^{-1} < 10$~Mpc, 
the BBN constraint obteined in this paper is the most stringent one for $0.1\,\text{Mpc}^{-1} \lesssim k \lesssim 4\times 10^8\, \text{Mpc}^{-1}$. 
Moreover, this constraint is valid even if the perturbations are compensated isocurvature perturbations because BBN can be affected only by the baryonic perturbations.

We have also applied the BBN constraint to the relaxation leptogenesis scenario where large baryon isocurvature perturbations are produced in the last $N_\text{last}$ $e$-fold of inflation.
It is found that the upper bound on $N_\text{last}$ is imposed as $N_\text{last} \lesssim 34$ for $T_R \lesssim 10^{12}$~GeV from the BBN constraint on baryonic isocurvature perturbations.

\section*{Acknowledgements}
This work was supported by JSPS KAKENHI Grant Nos. 17H01131 (M.K.) and 17K05434 (M.K.), MEXT KAKENHI Grant No. 15H05889 (M.K.), World Premier International Research Center Initiative (WPI Initiative), MEXT, Japan, Advanced Leading Graduate Course for Photon Science (K.I.), JSPS Research Fellowship for Young Scientists (K.I.), and the U.S. Department of Energy Grant No. DE\,-\,SC0009937 (A.K.).


\bibliographystyle{JHEP}

\providecommand{\href}[2]{#2}\begingroup\raggedright\endgroup

\end{document}